\documentclass[12pt]{article}

\usepackage{times}
\usepackage{rotating}
\usepackage[authoryear]{natbib}
\usepackage{amssymb,epsfig}
\usepackage[centertags]{amsmath}
\usepackage{bm}

\newcommand{\Es}{{\rm E}}
\newcommand{\Var}{{\rm var}}

\title{Testing hypotheses in the Birnbaum--Saunders distribution
under type-II censored samples}
\author{Artur J.~Lemonte,\quad Silvia L.~P.~Ferrari\\
{\small {\em Departamento de Estat\'istica, Universidade de S\~ao Paulo, Brazil}}}
\date{}

\begin{document}
\maketitle

\begin{abstract}

The two-parameter Birnbaum--Saunders distribution has been used
succesfully to model fatigue failure times.
Although censoring is typical in reliability and survival studies, little work has been published on the
analysis of censored data for this distribution. In this paper, we address the issue of performing testing
inference on the two parameters of the Birnbaum--Saunders distribution under
type-II right censored samples. The likelihood ratio statistic and a recently proposed
statistic, the gradient statistic, provide a convenient framework
for statistical inference in such a case, since they do not require to obtain, estimate or
invert an information matrix, which is an advantage in problems involving censored data.
An extensive Monte Carlo simulation study is carried out in order to investigate and compare
the finite sample performance of the likelihood ratio and the gradient tests.
Our numerical results show evidence that the gradient test should be preferred.
Three empirical applications are presented. \\

\end{abstract}

\section{Introduction}\label{introduction}

\cite{BSa1969a, BSa1969b} proposed a family of two-parameter distributions to model
failure time due to fatigue under cyclic loading and the assumption that failure
follows from the development and growth of a dominant crack.
This distribution is known as the two-parameter Birnbaum--Saunders ($\mathcal{BS}$)
distribution or as the fatigue life distribution.
The $\mathcal{BS}$ distribution is an attractive alternative to
the Weibull, gamma, and log-normal models, since its derivation 
considers the basic characteristics of the fatigue process.
A more general derivation was provided by \cite{Desmond1985}
based on a biological model and relaxing
several of the assumptions made by \cite{BSa1969a}. \cite{Desmond1986} investigated
the relationship between the $\mathcal{BS}$ distribution and the
inverse Gaussian distribution and demonstrated that the
$\mathcal{BS}$ distribution is an equal-weight mixture of
an inverse Gaussian distribution and its complementary reciprocal;
see also \cite{BhattacharyyaFries1982}.

\begin{figure}
\centering
\includegraphics[scale=0.42]{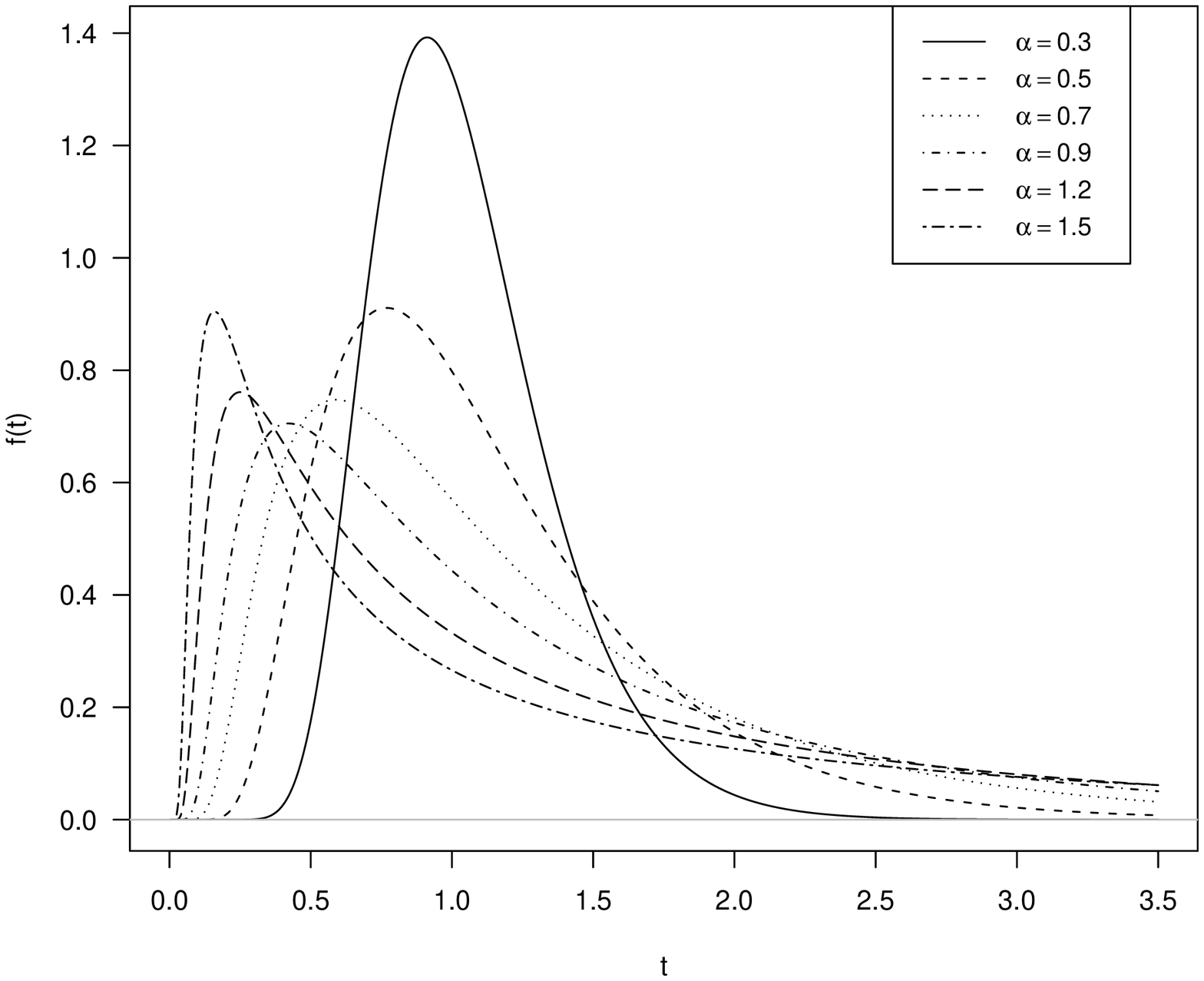}
\includegraphics[scale=0.42]{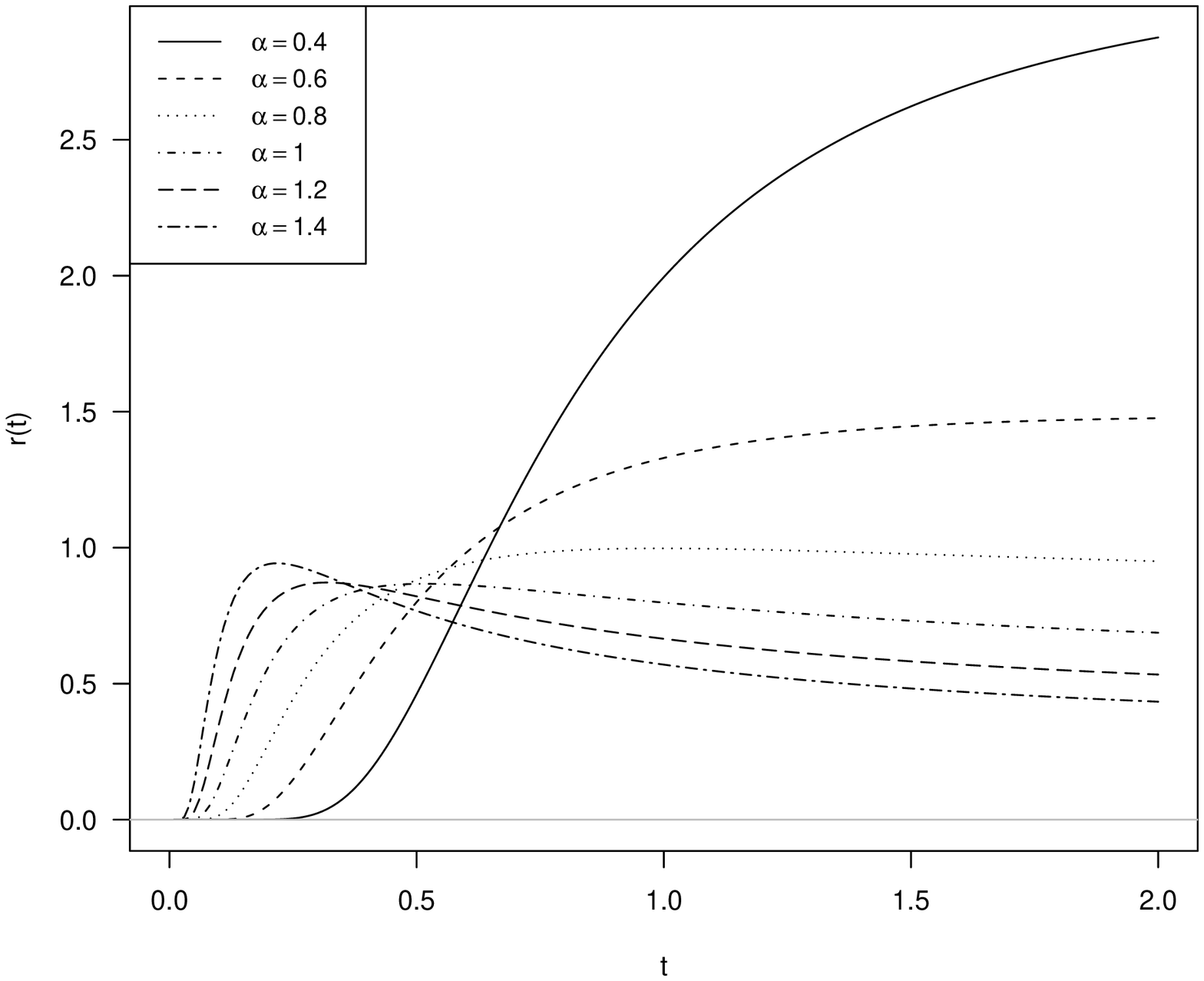}
\caption{Plots of the density and hazard ratio function; $\beta = 1$.}\label{fig1}
\end{figure}

The random variable $T$ is said to have a $\mathcal{BS}$ distribution with
parameters $\alpha, \beta > 0$, say $\mathcal{BS}(\alpha, \beta)$,
if its cumulative distribution function (cdf) is given by
\begin{equation}\label{cdf}
F(t)=\Phi(v), \qquad t > 0,
\end{equation}
where $\Phi(\cdot)$ is the standard normal distribution function,
$v=\rho(t/\beta)/\alpha$, $\rho(z)= z^{1/2}-z^{-1/2}$
and $\alpha$ and $\beta$ are shape and scale parameters,
respectively. Also, $\beta$  is the median of the distribution: $F(\beta) = \Phi(0) = 1/2$.
For any constant $k > 0$, it follows that $kT \sim\mathcal{BS}(\alpha, k\eta)$.
It is noteworthy that the reciprocal property holds for the $\mathcal{BS}$
distribution: $T^{-1} \sim\mathcal{BS}(\alpha, \beta^{-1})$; see
\cite{Saunders1974}. The probability density and hazard ratio functions
corresponding to (\ref{cdf}) are given by
\begin{equation}\label{pdf}
f(t)=\kappa(\alpha,\beta)t^{-3/2}(t+\beta)\exp\biggl\{-\frac{\tau(t/\beta)}{2\alpha^2}\biggr\},\qquad t>0,
\end{equation}
and
\[
r(t) = \frac{\kappa(\alpha,\beta)t^{-3/2}(t+\beta)\exp\{-\tau(t/\beta)/(2\alpha^2)\}}{1-\Phi(v)}, \qquad t>0,
\]
respectively, where $\kappa(\alpha,\beta)=\exp(\alpha^{-2})/(2\alpha\sqrt{2\pi\beta})$ and
$\tau(z)=z+z^{-1}$. The expected value, variance, skewness and
kurtosis are, respectively, $\Es(T) = \beta(1 + \alpha^2/2)$,
$\Var(T) = (\alpha\beta)^2(1 + 5\alpha^2/4)$,
$\mu_{3} = 16\alpha^2(11\alpha^2 + 6)/(5\alpha^2 + 4)^3$ and
$\mu_{4} = 3 + 6\alpha^2(93\alpha^2 + 41)/(5\alpha^2 + 4)^2$.
A general expression to obtain the moments of~(\ref{pdf}) was
derived by \cite{Rieck1999} and is given by
\[
\Es(T^{p})=\beta^p\biggl(\frac{K_{p+1/2}(\alpha^{-2})+K_{p-1/2}(\alpha^{-2})}{2K_{1/2}(\alpha^{-2})}\biggr),
\]
where $K_{\nu}(\cdot)$ denotes the modified Bessel function of the
third kind and order $\nu$. As noted before, if $T \sim\mathcal{BS}(\alpha, \beta)$, then
$T^{-1} \sim\mathcal{BS}(\alpha, \beta^{-1})$. It then follows that
$\Es(T^{-1}) = \beta^{-1}(1 + \alpha^2/2)$ and
$\Var(T^{-1}) = \alpha^2\beta^{-2}(1 + 5\alpha^2/4)$.
The $\mathcal{BS}$ distribution is positively skewed and
the asymmetry of the distribution decreases with $\alpha$.
Figure \ref{fig1} plots the density and hazard ratio function
for some values of $\alpha$ with $\beta = 1$. 
For applications of the $\mathcal{BS}$ distribution in reliability studies
see, for example, \cite{Balakrishnan-et-al-2007}, and for its use
outside this field see \cite{Leiva-et-al-2008} and \cite{Leiva-et-al-2009}.

The $\mathcal{BS}$ distribution has received significant attention
over the last few years. \cite{kundu-et-al-2008} discussed the shape of the hazard
function of the $\mathcal{BS}$ distribution. Improved frequentist
inference for the $\mathcal{BS}$ distribution is discussed in \cite{WuWong2004},
\cite{Cysneiros-et-al-2008} and \cite{LCNV07, LSCN08}, while estimation using reference
prior is presented in \cite{XiTang10}.
Extensions of the $\mathcal{BS}$ distribution are proposed in \cite{Diaz-Leiva05},
\cite{GPB09}, \cite{guiraud-et-al-2009} and \cite{Castillo-et-al-2009}.
Also, \cite{Bhatti2010} introduced the $\mathcal{BS}$ autoregressive conditional duration model and
\cite{Aslam-et-al-2010} developed various acceptance
sampling schemes based on the truncated life tests.

Little work has been published on the analysis of censored data for the
$\mathcal{BS}$ distribution although censoring is common in reliability 
and survival studies. \cite{Rieck1995} considered parametric
estimation for the $\mathcal{BS}$ distribution based on type-II
censored samples. \cite{Jeng2003} explored and compared different
procedures to compute confidence intervals for the parameters and
quantiles of the $\mathcal{BS}$ distribution for complete and type-I
censored data. \cite{Ng-et-al-2006} and \cite{Wang-et-al-2006}
discussed the maximum likelihood estimation of the parameters of
the $\mathcal{BS}$ distribution and
proposed a simple bias-reduction method to reduce the bias
of the maximum likelihood estimators (MLEs) based on type-II
censored samples and random censoring, respectively. Also,
\cite{FromLi2006} discussed the estimation of the parameters
of the $\mathcal{BS}$ distribution under censored samples.

A particularly useful censoring mechanism is the type-II (right) censoring,
which occurs when $n$ items are placed on test and the experiment is 
terminated when the first $m$ ($m<n$) items fail. Since life testing 
is often time consuming and expensive, type-II censoring may be used 
to reduce testing time and costs. This paper focus on testing inference 
for the $\mathcal{BS}$ distribution under type-II censored samples. 

A commonly used test procedure is the likelihood ratio test. 
A new criterion for testing hypotheses, referred to as the
{\it gradient test}, was proposed by \cite{Terrell2002}.
Its statistic shares the same first order asymptotic properties with the
likelihood ratio statistic. Like the likelihood ratio statistic and unlike the 
score and the Wald statistics, the gradient statistic
does not require computation of the information matrix, neither observed nor
expected, which is an advantage in problems involving censored samples.  
The chief goal of our paper is to present Monte Carlo simulations
in order to evaluate and compare the performance of the likelihood ratio and 
gradient tests
for the $\mathcal{BS}$ distribution under type-II right
censored samples. The other two tests will be briefly taken into account 
in our simulation study. To the best of our knowledge, there is no mention in the
statistical literature on the use of the gradient test
under censoring.

The paper is organized as follows. Section~\ref{tests} briefly describes
the  likelihood ratio and gradient tests. Section~\ref{bs_type-II} presents
inference on the parameters of the $\mathcal{BS}$ distribution
under type-II right censored samples.
Numerical results are presented and discussed in Section~\ref{simulations}.
We evaluate and compare the performance of the likelihood ratio and
gradient tests for testing hypotheses on the parameters of the
$\mathcal{BS}$ distribution under type-II right censored samples.
Applications are presented in Section~\ref{applications}.
Finally, Section~\ref{conclusions} closes the paper with some conclusions.

\section{Likelihood ratio and gradient tests}\label{tests}

Let $\ell(\bm{\theta})$ denote the total
log-likelihood function and consider the partition $\bm{\theta} =
(\bm{\theta}_{1}^{\top}, \bm{\theta}_{2}^{\top})^{\top}$,
where the dimensions of $\bm{\theta}_{1}$ and $\bm{\theta}_{2}$
are $q$ and $p-q$, respectively, i.e.~$\bm{\theta}$ is a $p$-vector of unknown parameters.
Let $\bm{U}_{\bm{\theta}}$ denote the score function for $\bm{\theta}$.
The partition for $\bm{\theta}$ induces the corresponding partition:
$\bm{U}_{\bm{\theta}} = (\bm{U}_{\bm{\theta}_1}^\top, \bm{U}_{\bm{\theta}_2}^\top)^\top$.
Suppose the interest lies in testing the composite null hypothesis
\[
\mathcal{H}_{0}:\bm{\theta}_{2} = \bm{\theta}_{20}
\]
against $\mathcal{H}_{1}:\bm{\theta}_{2}\neq\bm{\theta}_{20}$,
where $\bm{\theta}_{20}$ is a specified vector.
Hence, $\bm{\theta}_{1}$ is a vector of nuisance parameters.
The likelihood ratio ($LR$) and gradient ($S_T$) statistics for
testing $\mathcal{H}_{0}$ versus $\mathcal{H}_{1}$ are given, respectively, by
\[
LR = 2\bigl\{\ell(\widehat{\bm{\theta}}) - \ell(\widetilde{\bm{\theta}})\bigr\},
\qquad
S_{T} = \widetilde{\bm{U}}_{\bm{\theta}}^\top(\widehat{\bm{\theta}} - \widetilde{\bm{\theta}}),
\]
where $\widehat{\bm{\theta}}=(\widehat{\bm{\theta}}_1^\top,\widehat{\bm{\theta}}_2^\top)^\top$ and
$\widetilde{\bm{\theta}}=(\widetilde{\bm{\theta}}_1^\top,\bm{\theta}_{20}^{\top})^\top$
denote the MLEs of $\bm{\theta} = (\bm{\theta}_{1}^{\top}, \bm{\theta}_{2}^{\top})^{\top}$ under
$\mathcal{H}_{1}$ and $\mathcal{H}_{0}$, respectively, and
$\widetilde{\bm{U}}_{\bm{\theta}} = \bm{U}_{\bm{\theta}}(\widetilde{\bm{\theta}})
= (\widetilde{\bm{U}}_{\bm{\theta}_{1}}^\top,\widetilde{\bm{U}}_{\bm{\theta}_{2}}^\top)^\top$,
with $\widetilde{\bm{U}}_{\bm{\theta}_1} = \bm{U}_{\bm{\theta}_1}(\widetilde{\bm{\theta}})$
and $\widetilde{\bm{U}}_{\bm{\theta}_2} = \bm{U}_{\bm{\theta}_2}(\widetilde{\bm{\theta}})$.
Since $\widetilde{\bm{U}}_{\bm{\theta}_1} = \bm{0}$, the gradient statistic can be written as
$S_{T} = \widetilde{\bm{U}}_{\bm{\theta}_2}^\top(\widehat{\bm{\theta}}_{2} - \bm{\theta}_{20})$.
The limiting distribution of $LR$ and $S_T$ is $\chi_{p-q}^2$ under
$\mathcal{H}_{0}$ and $\chi_{p-q,\lambda}^{2}$, i.e.~a noncentral chi-square
distribution with $p-q$ degrees of freedom and an appropriate noncentrality parameter $\lambda$,
under $\mathcal{H}_{1}$. The null hypothesis is rejected for a given nominal level, $\gamma$ say,
if the test statistic exceeds the upper $100(1-\gamma)\%$ quantile of the $\chi_{p-q}^{2}$
distribution.

\cite{Terrell2002} points out that the
gradient statistic ``is not transparently non-negative, even
though it must be so asymptotically.'' His Theorem 2 implies that
if the log-likelihood function is concave and is differentiable at
$\widetilde{\bm{\theta}}$, then $S_{T}\ge 0$. Recently, \cite{LemonteFerrari2010a} derived
the nonnull asymptotic expansion of the distribution of the gradient test statistic
for a composite hypothesis under a sequence of local alternative
hypotheses converging to the null hypothesis at rate $n^{-1/2}$.
The authors also compared the local powers of the
gradient, likelihood ratio, Wald and score tests.
In general, no test is uniformly better than the
others as far as local power is concerned. \cite{LemonteFerrari2010b}
obtained the nonnull asymptotic expansions of the
likelihood ratio, Wald, score and gradient statistics in 
Birnbaum--Saunders regression models when no censoring is present.
An interesting finding is that, up to an error of order $n^{-1}$, 
the four tests have the same local power in this class of models.
Their simulation study evidenced that the score and the gradient tests
perform better than the likelihood ratio and Wald tests in small and
moderate-sized samples.

\section{Inference under type-II censored samples}\label{bs_type-II}

Let $\{t_{1},\ldots,t_{m}\}$ be an ordered type-II right censored random sample obtained
from $n$ units placed on a life-testing experiment wherein each unit has its lifetime
following the $\mathcal{BS}$ distribution, with the largest $(n-m)$ lifetimes having been
censored. Let $\bm{\theta} = (\alpha, \beta)^\top$ be the parameter vector.
For notational convenience, let
\begin{equation*}\label{quantities}
H(z) = \frac{\phi(z)}{1 - \Phi(z)}, \qquad
h(z) = \frac{\tau(z^{1/2})}{\alpha}, \qquad
t_{i}^* = \frac{t_{i}}{\beta},
\end{equation*}
where $\phi(\cdot)$ is the standard normal density function.

The likelihood function can be written as \citep{BalaCohen1991}
\[
L(\bm{\theta}) = \Biggl\{\prod_{i=1}^{m}\kappa(\alpha,\beta)t_{i}^{-3/2}
(t_{i}+\beta)\exp\biggl(-\frac{\tau(t_{i}^*)}{2\alpha^2}\biggr)\Biggr\}
\frac{n!\{1 - \Phi(v_{m})\}^{n-m}}{m!(n-m)!},
\]
where $v_{m} = \rho(t_{m}^*)/\alpha$.
Thus, the log-likelihood function, except for a constant term, is given by
\begin{align*}\label{loglik-type-II}
\ell(\bm{\theta}) = \ell(\alpha, \beta) &= m\log\{\kappa(\alpha,\beta)\}
+\sum_{i=1}^{m}\log(t_{i}+\beta)-\frac{1}{2\alpha^2}\sum_{i=1}^m\tau(t_{i}^*)\\
&\quad+ (n - m)\log\{1 - \Phi(v_{m})\}.
\end{align*}
By taking partial derivatives of the log-likelihood function with respect to
$\alpha$ and $\beta$ we obtain the components of the score vector,
$\bm{U}_{\bm{\theta}} = (U_{\alpha}, U_{\beta})^\top$:
\[
U_\alpha = -\frac{m}{\alpha}\biggl(1 + \frac{2}{\alpha^2}\biggr)
+ \frac{1}{\alpha^3}\sum_{i=1}^{m}\tau(t_{i}^*)
+ \frac{(n-m)v_{m}H(v_{m})}{\alpha},
\]
\[
U_{\beta} = -\frac{m}{2\beta} + \sum_{i=1}^{m}\frac{1}{t_{i} + \beta}
 + \frac{1}{2\alpha^2\beta}\sum_{i=1}^{m}\rho(t_{i}^{*2})
+ \frac{(n-m)h(t_{m}^*)H(v_{m})}{2\beta}.
\]
Setting $U_{\alpha}$ and $U_{\beta}$ equal to zero yields the MLE
$\widehat{\bm{\theta}} = (\widehat{\alpha},\widehat{\beta})^\top$
of $\bm{\theta} = (\alpha,\beta)^\top$. These equations cannot be solved
analytically and statistical software can be used to solve them numerically
via iterative methods. The BFGS method \citep[see, for example,][]{NocedalWright1999,
Press-et-al-2007} with analytical derivatives has been used
for maximizing the log-likelihood function $\ell(\bm{\theta})$.

We now consider hypothesis testing on the parameters $\alpha$ and $\beta$ based on
the likelihood ratio and gradient statistics.
The interest lies in testing the null hypotheses
\begin{equation*}\label{hypotheses}
\mathcal{H}_{00}:\alpha = \alpha^{(0)}, \qquad
\mathcal{H}_{01}:\beta = \beta^{(0)},
\end{equation*}
which are tested against $\mathcal{H}_{10}:\alpha\neq \alpha^{(0)}$ and
$\mathcal{H}_{11}:\beta\neq \beta^{(0)}$, respectively. Here $\alpha^{(0)}$
and $\beta^{(0)}$ are positive known scalars. For testing
$\mathcal{H}_{00}$, the likelihood ratio  and
gradient statistics are given by
\[
LR_{(\alpha)} = 2\bigl\{\ell(\widehat{\beta},\widehat{\alpha})
- \ell(\alpha^{(0)},\widetilde{\beta})\bigr\},
\qquad
S_{T(\alpha)} = \widetilde{U}_{\alpha}(\widehat{\alpha} - \alpha^{(0)}),
\]
where $\widehat{\alpha}$ and $\widehat{\beta}$ are the unrestricted MLEs
obtained from the maximization of $\ell(\bm{\theta})$ under the alternative hypotheses,
$\widetilde{\beta}$ is the restricted
MLE obtained from the maximization of $\ell(\bm{\theta})$ under $\mathcal{H}_{00}$ and
$\widetilde{U}_{\alpha} = U_{\alpha}(\alpha^{(0)},\widetilde{\beta})$. For
testing $\mathcal{H}_{01}$, we have
\[
LR_{(\beta)} = 2\bigl\{\ell(\widehat{\beta},\widehat{\alpha})
- \ell(\widetilde{\alpha},\beta^{(0)})\bigr\},
\qquad
S_{T(\beta)} = \widetilde{U}_{\beta}(\widehat{\beta} - \beta^{(0)}).
\]
Here, $\widetilde{\alpha}$ is the restricted
MLE obtained from the maximization of $\ell(\bm{\theta})$ under
the hypothesis $\mathcal{H}_{01}$
and $\widetilde{U}_{\beta} = U_{\beta}(\widetilde{\alpha},\beta^{(0)})$.
The limiting distribution of these statistics is $\chi_{1}^2$ under
the respective null hypothesis. In both cases, the null hypothesis is rejected if the chosen test
statistic exceeds the upper $100(1-\gamma)\%$ quantile of the $\chi_{1}^{2}$
distribution.

Theorem 3 in \cite{Terrell2002} points out an important feature of the
gradient test. It suggests that we can, in general, improve the approximation
of the distribution of the gradient statistic by a chi-square distribution
under the null hypothesis by using a less biased estimator.
On the other hand, Theorem 2 in \cite{Terrell2002} indicates that if the estimate used
in the gradient statistic is not the maximum likelihood,
the non-negativity of the gradient statistic is not guaranteed.

A bias-corrected estimator for $\alpha$ was derived in \cite{Ng-et-al-2006} and is given by
\[
\bar{\alpha} = \widehat{\alpha}
\biggl\{1 - \frac{1}{n}\biggl[1 + 2.5\biggl(1 - \frac{m}{n}\biggr)\biggr]\biggr\}^{-1}.
\]
The authors showed through Monte Carlo simulations that
$\bar{\alpha}$ is less biased than the
original MLE, $\widehat{\alpha}$, of $\alpha$.
Therefore, based on this estimator, we can define the adjusted gradient statistic
$S_{T(\alpha)}^{*} = \widetilde{U}_{\alpha}(\bar{\alpha} - \alpha^{(0)})$
for testing the null hypothesis $\mathcal{H}_{00}:\alpha = \alpha^{(0)}$.
The limiting distribution of $S_{T(\alpha)}^{*}$ is also $\chi_{1}^2$ under
$\mathcal{H}_{00}$. It is expected that the gradient test
that uses the statistic $S_{T(\alpha)}^{*}$ have
better size performance than the gradient test based on the original gradient statistic
$S_{T(\alpha)}$. A disadvantage of the adjusted gradient statistic
$S_{T(\alpha)}^{*}$ is that it can take on negative values. However,
in order to avoid negative values we shall redefine $S_{T(\alpha)}^{*}$ as
\[
S_{T(\alpha)}^{*} = \max\{0, \widetilde{U}_{\alpha}(\bar{\alpha} - \alpha^{(0)})\}.
\]
An adjusted gradient statistic for testing
the null hypothesis $\mathcal{H}_{01}:\beta = \beta^{(0)}$
is not considered here because a bias-corrected estimator
for $\beta$ is not available. In the next section, we
shall present an extensive Monte Carlo study in order to
evaluate and compare the performance of the tests presented in this section.

\section{Numerical results}\label{simulations}

In this section we shall present an extensive Monte Carlo simulation study
in which we evaluate and compare the finite sample performance
of the likelihood ratio and gradient tests for
testing hypotheses on the parameters of the
$\mathcal{BS}$ distribution under type-II right censored samples.
We set the degree of censoring (d.o.c.) at
$0(10)50\%$, the sample size at $n = 20$ and 40,
and the shape parameter at $\alpha = 0.1, 0.3, 0.5, 0.75$ and 1.0.
Without loss of generality, the scale parameter $\beta$ was kept fixed at 1.0.
The nominal levels of the tests were $\gamma = 10\%$
and 5\%. The number of Monte Carlo replications
was 10000. All the Monte Carlo simulation experiments were
performed using the {\sf Ox} matrix programming language \citep{DcK2006}.
{\sf Ox} is freely distributed for academic
purposes and available at http://www.doornik.com.

Table \ref{tab1_new} presents the null rejection rates
(entries are percentages) of the tests which are based on the statistics
$LR_{(\alpha)}$, $S_{T(\alpha)}$ and $S_{T(\alpha)}^*$ for testing the null hypothesis
$\mathcal{H}_{00}:\alpha=\alpha^{(0)}$.
Our main findings are as follows.
First, for complete data without censoring (d.o.c.~= 0\%), the tests based on
$S_{T(\alpha)}$ and $S_{T(\alpha)}^*$ are less size distorted
than the test that uses $LR_{(\alpha)}$; in fact, they
produce null rejection rates that are very close to the nominal
levels in all the cases considered. In addition, the adjusted gradient test
is slightly superior than the original gradient test.
For example, for $n=20$, $\alpha^{(0)} = 0.5$ and $\gamma = 10\%$,
the null rejection rates are 11.48\% ($LR_{(\alpha)}$),
9.65\% ($S_{T(\alpha)}$) and 9.81\% ($S_{T(\alpha)}^*$). Second,
the size distortion of all tests increases with the d.o.c., the
likelihood ratio test and the original gradient test
displaying null rejection rates that
are, respectively, greater and smaller than the nominal level.
The adjusted gradient test is the less size distorted in most of the cases.
For instance, for $n=20$, $\alpha^{(0)} = 0.75$ and $\gamma = 5\%$,
the null rejection rates are 6.17\% ($LR_{(\alpha)}$),
4.28\% ($S_{T(\alpha)}$) and 4.77\% ($S_{T(\alpha)}^*$) for d.o.c.~= 10\%,
and 7.22\% ($LR_{(\alpha)}$), 4.15\% ($S_{T(\alpha)}$)
and 4.96\% ($S_{T(\alpha)}^*$) for d.o.c.~= 30\%.
Note that all the tests become less size
distorted as the sample size increases, as expected.
\begin{table}[!htp]
\begin{center}
{\footnotesize
\caption{Null rejection rates (\%) for $\mathcal{H}_{00}:\alpha=\alpha^{(0)}$.}\label{tab1_new}
\begin{tabular}{cl cccccc c cccccc}\hline
      &  & \multicolumn{6}{c}{$n=20$} && \multicolumn{6}{c}{$n=40$}\\ \cline{3-8} \cline{10-15}
d.o.c.&  & \multicolumn{2}{c}{$LR_{(\alpha)}$} & \multicolumn{2}{c}{$S_{T(\alpha)}$} & \multicolumn{2}{c}{$S_{T(\alpha)}^{*}$}
        && \multicolumn{2}{c}{$LR_{(\alpha)}$} & \multicolumn{2}{c}{$S_{T(\alpha)}$} & \multicolumn{2}{c}{$S_{T(\alpha)}^{*}$}\\
 (\%) & $\alpha^{(0)}$ & 10\%  & 5\% & 10\% & 5\%  & 10\%  & 5\% && 10\%  & 5\% & 10\% & 5\%  & 10\%  & 5\% \\\cline{1-8} \cline{10-15}
0 & 0.1 & 11.53 & 6.07 & 9.70 & 4.25 &  9.83 & 4.52 && 10.76 & 5.23 &  9.67 & 4.82 & 9.81 & 5.16 \\
  & 0.3 & 11.50 & 6.10 & 9.70 & 4.22 &  9.78 & 4.48 && 10.74 & 5.22 &  9.69 & 4.83 & 9.83 & 5.15 \\
  & 0.5 & 11.48 & 6.15 & 9.65 & 4.21 &  9.81 & 4.45 && 10.75 & 5.24 &  9.71 & 4.80 & 9.85 & 5.12 \\
  & 0.75& 11.55 & 6.24 & 9.58 & 4.22 &  9.78 & 4.41 && 10.77 & 5.29 &  9.74 & 4.79 & 9.83 & 5.09 \\
  & 1.0 & 11.73 & 6.25 & 9.62 & 4.26 &  9.78 & 4.46 && 10.84 & 5.36 &  9.67 & 4.81 & 9.79 & 5.10 \\ \cline{1-8} \cline{10-15}

10& 0.1 & 11.85 & 6.12 & 9.57 & 4.29 &  9.84 & 4.90 && 10.79 & 5.70 &  9.63 & 4.61 & 9.87 & 4.79 \\
  & 0.3 & 11.86 & 6.12 & 9.55 & 4.29 &  9.82 & 4.88 && 10.77 & 5.70 &  9.62 & 4.60 & 9.85 & 4.75 \\
  & 0.5 & 11.85 & 6.15 & 9.59 & 4.29 &  9.81 & 4.85 && 10.78 & 5.67 &  9.63 & 4.56 & 9.80 & 4.73 \\
  & 0.75& 11.96 & 6.17 & 9.59 & 4.28 &  9.82 & 4.77 && 10.76 & 5.66 &  9.59 & 4.51 & 9.88 & 4.74 \\
  & 1.0 & 12.09 & 6.19 & 9.58 & 4.35 &  9.77 & 4.73 && 10.90 & 5.69 &  9.67 & 4.51 & 9.89 & 4.73 \\ \cline{1-8} \cline{10-15}

20& 0.1 & 12.26 & 6.50 & 9.80 & 4.10 & 10.01 & 5.09 && 10.71 & 5.30 &  9.23 & 4.36 & 9.38 & 4.63 \\
  & 0.3 & 12.24 & 6.52 & 9.74 & 4.12 &  9.96 & 5.06 && 10.77 & 5.26 &  9.24 & 4.36 & 9.29 & 4.63 \\
  & 0.5 & 12.27 & 6.50 & 9.79 & 4.16 &  9.94 & 5.04 && 10.78 & 5.25 &  9.29 & 4.37 & 9.19 & 4.63 \\
  & 0.75& 12.26 & 6.55 & 9.68 & 4.20 &  9.91 & 5.00 && 10.82 & 5.36 &  9.32 & 4.27 & 9.33 & 4.55 \\
  & 1.0 & 12.27 & 6.64 & 9.72 & 4.22 &  9.81 & 4.93 && 10.85 & 5.36 &  9.26 & 4.28 & 9.21 & 4.52 \\ \cline{1-8} \cline{10-15}

30& 0.1 & 12.98 & 6.98 & 9.76 & 4.30 & 10.30 & 5.12 && 11.10 & 5.55 &  9.40 & 4.43 & 9.72 & 4.64 \\
  & 0.3 & 13.04 & 6.96 & 9.74 & 4.26 & 10.17 & 5.10 && 11.11 & 5.52 &  9.40 & 4.33 & 9.69 & 4.67 \\
  & 0.5 & 13.04 & 7.06 & 9.82 & 4.28 & 10.14 & 5.04 && 10.98 & 5.50 &  9.38 & 4.30 & 9.66 & 4.73 \\
  & 0.75& 13.04 & 7.22 & 9.81 & 4.15 &  9.96 & 4.96 && 10.97 & 5.58 &  9.34 & 4.24 & 9.56 & 4.63 \\
  & 1.0 & 13.08 & 7.26 & 9.76 & 4.19 &  9.76 & 4.91 && 10.90 & 5.59 &  9.24 & 4.24 & 9.27 & 4.60 \\ \cline{1-8} \cline{10-15}

40& 0.1 & 13.30 & 7.23 & 9.42 & 4.27 &  9.93 & 5.15 && 11.78 & 5.89 &  9.96 & 4.79 & 9.93 & 5.38 \\
  & 0.3 & 13.31 & 7.25 & 9.40 & 4.21 &  9.95 & 5.18 && 11.75 & 5.89 &  9.98 & 4.81 & 9.95 & 5.36 \\
  & 0.5 & 13.29 & 7.12 & 9.42 & 4.14 &  9.92 & 5.13 && 11.74 & 5.93 & 10.00 & 4.75 & 9.88 & 5.26 \\
  & 0.75& 13.25 & 7.26 & 9.32 & 4.05 &  9.90 & 5.04 && 11.73 & 5.87 &  9.83 & 4.78 & 9.82 & 5.24 \\
  & 1.0 & 13.52 & 7.32 & 9.20 & 3.95 &  9.73 & 4.95 && 11.73 & 5.95 &  9.73 & 4.68 & 9.74 & 5.09 \\ \cline{1-8} \cline{10-15}

50& 0.1 & 14.35 & 7.99 & 9.30 & 3.95 &  9.78 & 4.91 && 11.42 & 6.02 &  9.38 & 4.57 & 9.61 & 5.10 \\
  & 0.3 & 14.37 & 7.94 & 9.24 & 3.88 &  9.70 & 4.84 && 11.39 & 6.05 &  9.42 & 4.61 & 9.64 & 5.07 \\
  & 0.5 & 14.18 & 7.92 & 9.20 & 3.89 &  9.65 & 4.85 && 11.39 & 6.12 &  9.33 & 4.55 & 9.57 & 5.03 \\
  & 0.75& 14.37 & 8.01 & 9.04 & 3.81 &  9.56 & 4.71 && 11.42 & 6.17 &  9.39 & 4.45 & 9.39 & 5.02 \\
  & 1.0 & 14.60 & 8.05 & 8.67 & 3.58 &  9.23 & 4.59 && 11.51 & 6.23 &  9.12 & 4.33 & 9.22 & 4.91 \\ \hline
\end{tabular}
}
\end{center}
\end{table}

The null rejection rates of the tests which are
based on $LR_{(\beta)}$ and $S_{T(\beta)}$
for testing the null hypothesis $\mathcal{H}_{01}:\beta=1$
are presented in Table~\ref{tab3_new}.
For complete data without censoring (d.o.c. = 0\%), the
gradient test is less size distorted than the likelihood test
in all the cases. For example, for $n=20$, $\alpha = 0.5$ and $\gamma = 10\%$,
the null rejection rates are 11.98\% ($LR_{(\beta)}$) and 10.53\% ($S_{T(\beta)}$).
Additionally, the size distortion of the tests increases with the
degree of censoring and, similarly to what occurs for testing on the parameter $\alpha$,
the likelihood ratio test presented a liberal behavior,
while the gradient test was conservative in the majority of the cases.
For instance, when $n=20$, $\alpha=0.5$ and $\gamma=10\%$, the null rejection rates
are 12.28\% ($LR_{(\beta)}$) and 9.88\% ($S_{T(\beta)}$) for d.o.c. = 20\% and
13.57\% ($LR_{(\beta)}$) and 9.47\% ($S_{T(\beta)}$) for d.o.c. = 50\%.
It should be noticed that the gradient test is less size
distorted than the likelihood test in all the cases considered.
Again, the tests become less size distorted as the sample size increases.
\begin{table}[!htp]
\begin{center}
{\footnotesize
\caption{Null rejection rates (\%) for $\mathcal{H}_{01}:\beta=1$.}\label{tab3_new}
\begin{tabular}{cl cccc c cccc}\hline
      &  & \multicolumn{4}{c}{$n=20$} &&  \multicolumn{4}{c}{$n=40$}\\ \cline{3-6}\cline{8-11}
d.o.c.&  & \multicolumn{2}{c}{$LR_{(\beta)}$} & \multicolumn{2}{c}{$S_{T(\beta)}$}
         && \multicolumn{2}{c}{$LR_{(\beta)}$} & \multicolumn{2}{c}{$S_{T(\beta)}$}\\
 (\%) & $\alpha$ & 10\% & 5\% & 10\%  & 5\% && 10\%  & 5\% & 10\% & 5\% \\\cline{1-6}\cline{8-11}
0 & 0.1 & 12.06 & 6.43& 10.74 & 4.98 && 11.00 & 5.43& 10.38 & 4.92 \\
  & 0.3 & 11.95 & 6.44& 10.57 & 5.03 && 11.02 & 5.44& 10.37 & 4.94 \\
  & 0.5 & 11.98 & 6.52& 10.53 & 4.94 && 11.00 & 5.50& 10.40 & 4.95 \\
  & 0.75& 11.78 & 6.49& 10.33 & 4.89 && 11.14 & 5.56& 10.26 & 5.08 \\
  & 1.0 & 11.74 & 6.74& 10.21 & 4.93 && 11.23 & 5.81& 10.33 & 5.08 \\ \cline{1-6}\cline{8-11}

10& 0.1 & 12.00 & 6.44& 10.69 & 4.70 && 10.82 & 5.59& 10.17 & 4.76 \\
  & 0.3 & 12.01 & 6.40& 10.40 & 4.70 && 10.80 & 5.51& 10.16 & 4.82 \\
  & 0.5 & 11.96 & 6.40& 10.22 & 4.63 && 10.82 & 5.45& 10.18 & 4.81 \\
  & 0.75& 11.91 & 6.27&  9.89 & 4.51 && 11.02 & 5.51& 10.15 & 4.83 \\
  & 1.0 & 11.87 & 6.29&  9.78 & 4.60 && 11.09 & 5.64& 10.05 & 4.80 \\ \cline{1-6}\cline{8-11}

20& 0.1 & 12.24 & 6.44& 10.08 & 4.57 && 11.13 & 5.55& 10.07 & 4.75 \\
  & 0.3 & 12.31 & 6.41& 10.00 & 4.68 && 11.14 & 5.52&  9.96 & 4.65 \\
  & 0.5 & 12.28 & 6.36&  9.88 & 4.52 && 11.00 & 5.55&  9.96 & 4.56 \\
  & 0.75& 12.32 & 6.39&  9.72 & 4.59 && 11.01 & 5.59&  9.81 & 4.58 \\
  & 1.0 & 12.32 & 6.53&  9.46 & 4.44 && 11.16 & 5.47&  9.80 & 4.61 \\ \cline{1-6}\cline{8-11}

30& 0.1 & 12.87 & 6.95& 10.53 & 4.66 && 11.17 & 5.60&  9.97 & 4.55 \\
  & 0.3 & 12.92 & 6.97& 10.23 & 4.50 && 11.18 & 5.62&  9.95 & 4.63 \\
  & 0.5 & 12.99 & 6.99&  9.98 & 4.39 && 11.11 & 5.63&  9.81 & 4.57 \\
  & 0.75& 13.01 & 7.08&  9.70 & 4.31 && 11.01 & 5.59&  9.65 & 4.47 \\
  & 1.0 & 13.12 & 6.91&  9.41 & 4.39 && 11.07 & 5.69&  9.36 & 4.45 \\ \cline{1-6}\cline{8-11}

40& 0.1 & 13.14 & 7.08& 10.11 & 4.38 && 11.32 & 5.84&  9.90 & 4.45 \\
  & 0.3 & 13.14 & 7.05&  9.67 & 4.38 && 11.32 & 5.85&  9.65 & 4.40 \\
  & 0.5 & 13.12 & 7.05&  9.40 & 4.26 && 11.31 & 5.85&  9.59 & 4.34 \\
  & 0.75& 13.10 & 7.03&  9.05 & 4.29 && 11.22 & 5.90&  9.52 & 4.41 \\
  & 1.0 & 13.14 & 6.99&  8.80 & 4.39 && 11.17 & 5.92&  9.34 & 4.24 \\ \cline{1-6}\cline{8-11}

50& 0.1 & 13.49 & 7.85&  9.90 & 4.22 && 11.70 & 5.98& 10.08 & 4.49 \\
  & 0.3 & 13.52 & 7.82&  9.67 & 4.06 && 11.71 & 5.98&  9.88 & 4.42 \\
  & 0.5 & 13.57 & 7.75&  9.47 & 4.24 && 11.71 & 5.97&  9.70 & 4.41 \\
  & 0.75& 13.54 & 7.73&  8.81 & 4.06 && 11.75 & 6.02&  9.38 & 4.34 \\
  & 1.0 & 13.58 & 7.69&  8.19 & 4.23 && 11.74 & 6.03&  9.02 & 4.33 \\ \hline
\end{tabular}
}
\end{center}
\end{table}

Table~\ref{tab5_new} contains the nonnull rejection rates (powers) of the tests.
We set $\alpha = 0.5$, $\beta = 1$, $n = 80, 120$ and 150,
and the degree of censoring at 0\%, 20\% and 40\%.
The rejection rates were obtained under the alternative hypotheses
$\mathcal{H}_{10}:\alpha = \delta_{1}$ and $\mathcal{H}_{11}:\beta = \delta_{2}$,
for different values of $\delta_{1}$ and $\delta_{2}$.
The test that uses $LR_{(\alpha)}$ presented smaller powers for testing
hypotheses on the parameter $\alpha$ than the tests
which are based on the statistics $S_{T(\alpha)}$ and $S_{T(\alpha)}^*$.
For example, when $n=80$, d.o.c.~= 20\% and $\delta_{1} = 0.60$,
the nonnull rejection rates are 58.92\% ($LR_{(\alpha)}$),
62.02\% ($S_{T(\alpha)}$) and 65.53\% ($S_{T(\alpha)}^*$).
Additionally, the tests which use $LR_{(\beta)}$ and
$S_{T(\beta)}$ for testing hypothesis on the parameter $\beta$ have similar powers,
the gradient test being slightly more powerful than the likelihood test.
For instance, when $n=150$, d.o.c.~= 0\% and $\delta_{2} = 1.13$,
the nonnull rejection rates are 92.02\% ($LR_{(\beta)}$) and 92.33\% ($S_{T(\beta)}$).
Note that the powers of the all the tests decrease as the
degree of censoring increases. We also note that the powers of the tests increase with $n$
and also with $\delta_{1}$ and $\delta_{2}$, as expected.
\begin{table}[!htp]
\begin{center}
{\footnotesize
\caption{Nonnull rejection rates (\%): $\alpha = 0.5$,  $\beta = 1$,
$\gamma = 10\%$ and different sample sizes.}\label{tab5_new}
\begin{tabular}{cc cccc c ccc}\hline
    &    d.o.c.      & \multicolumn{4}{c}{$\mathcal{H}_{10}:\alpha = \delta_{1}$}
    &                & \multicolumn{3}{c}{$\mathcal{H}_{11}:\beta = \delta_{2}$}\\ \cline{3-6}\cline{8-10}
$n$ & (\%) &  $\delta_{1}$ & $LR_{(\alpha)}$ & $S_{T(\alpha)}$ & $S_{T(\alpha)}^*$
           && $\delta_{2}$ & $LR_{(\beta)}$ & $S_{T(\beta)}$\\ \cline{1-6}\cline{8-10}
80 & 0  & 0.50 & 10.71& 10.11& 10.25 && 1.00 & 10.53& 10.20 \\
   &    & 0.54 & 23.90& 25.84& 27.97 && 1.04 & 19.06& 19.50 \\
   &    & 0.58 & 56.16& 58.99& 62.18 && 1.10 & 54.60& 55.66 \\
   &    & 0.60 & 71.59& 73.97& 76.45 && 1.13 & 73.29& 73.96 \\\cline{2-6}\cline{8-10}

   & 20 & 0.50 & 10.65&  9.55&  9.68 && 1.00 & 10.55& 10.10 \\
   &    & 0.54 & 19.34& 20.93& 23.62 && 1.04 & 18.67& 19.27 \\
   &    & 0.58 & 44.89& 48.51& 52.02 && 1.10 & 52.70& 54.15 \\
   &    & 0.60 & 58.92& 62.02& 65.53 && 1.13 & 70.86& 72.06 \\\cline{2-6}\cline{8-10}

   & 40 & 0.50 & 10.93&  9.92&  9.72 && 1.00 & 10.30&  9.51 \\
   &    & 0.54 & 15.74& 17.20& 19.83 && 1.04 & 17.34& 18.47 \\
   &    & 0.58 & 33.78& 37.30& 40.92 && 1.10 & 47.69& 50.16 \\
   &    & 0.60 & 44.65& 48.93& 52.86 && 1.13 & 64.54& 67.06 \\\cline{1-6}\cline{8-10}

120& 0  & 0.50 & 10.53& 10.17& 10.18 && 1.00 & 10.62& 10.37 \\
   &    & 0.54 & 30.34& 32.59& 34.90 && 1.04 & 22.74& 23.25 \\
   &    & 0.58 & 72.73& 74.76& 76.52 && 1.10 & 69.69& 70.35 \\
   &    & 0.60 & 86.61& 87.73& 88.78 && 1.13 & 87.01& 87.47 \\\cline{2-6}\cline{8-10}

   & 20 & 0.50 & 10.64& 10.02& 10.14 && 1.00 & 10.19&  9.80 \\
   &    & 0.54 & 24.58& 26.43& 28.95 && 1.04 & 22.78& 23.46 \\
   &    & 0.58 & 60.06& 62.86& 65.65 && 1.10 & 68.10& 69.23 \\
   &    & 0.60 & 75.07& 77.04& 79.19 && 1.13 & 84.79& 85.37 \\\cline{2-6}\cline{8-10}

   & 40 & 0.50 & 10.65&  9.90&  9.81 && 1.00 & 10.21&  9.65 \\
   &    & 0.54 & 19.02& 21.07& 23.71 && 1.04 & 20.63& 22.17 \\
   &    & 0.58 & 45.24& 48.49& 52.05 && 1.10 & 60.84& 62.71 \\
   &    & 0.60 & 59.85& 62.66& 65.77 && 1.13 & 79.08& 80.55 \\\cline{1-6}\cline{8-10}

150& 0  & 0.50 &  9.98&  9.72&  9.76 && 1.00 & 10.15& 10.00 \\
   &    & 0.54 & 35.53& 38.04& 40.24 && 1.04 & 26.34& 26.97 \\
   &    & 0.58 & 80.99& 82.36& 83.63 && 1.10 & 77.31& 77.87 \\
   &    & 0.60 & 92.42& 93.24& 93.82 && 1.13 & 92.02& 92.33 \\\cline{2-6}\cline{8-10}

   & 20 & 0.50 & 10.13&  9.94&  9.99 && 1.00 & 10.39& 10.15 \\
   &    & 0.54 & 28.30& 30.38& 33.01 && 1.04 & 25.52& 26.39 \\
   &    & 0.58 & 68.24& 70.56& 73.17 && 1.10 & 75.83& 76.92 \\
   &    & 0.60 & 83.46& 84.88& 86.39 && 1.13 & 91.42& 91.90 \\\cline{2-6}\cline{8-10}

   & 40 & 0.50 & 10.51& 10.06&  9.85 && 1.00 & 10.33& 10.02 \\
   &    & 0.54 & 21.85& 23.99& 26.53 && 1.04 & 22.49& 23.86 \\
   &    & 0.58 & 53.25& 56.23& 59.28 && 1.10 & 68.41& 70.32 \\
   &    & 0.60 & 68.45& 70.74& 73.44 && 1.13 & 86.73& 87.78 \\\hline
\end{tabular}
}
\end{center}
\end{table}

A comment on the values of the shape parameter chosen for the simulation
study, namely $\alpha=0.1, 0.3, 0.5, 0.75$ and 1.0, is in order.
Following \cite{DupuisMills1998}, in most practical applications, $0<\alpha\leq 1$;
see, for instance, the applications considered in \cite{Sanhueza-et-al-2008} and
\cite{Leiva-et-al-2008}. As showed in \citet[][p.~46--48]{HoylandRausand1994}, this
follows from the derivation of the $\mathcal{BS}$ distribution.
As a result, most of the simulation experiments for the $\mathcal{BS}$ distribution
have been conducted for values of $\alpha$ between 0 and 1; see, for example,
\cite{DupuisMills1998}, \cite{WuWong2004}, \cite{Ng-et-al-2006},
\cite{Wang-et-al-2006}, \cite{LCNV07, LSCN08} and \cite{Cysneiros-et-al-2008}.
Likewise, we chose values of $\alpha$ in the unit interval to run our simulations and to analyse the simulation
findings in detail. We now briefly comment on simulation results for larger values of $\alpha$.
We conducted Monte Carlo experiments for various values of $\alpha>1$. In general,
for testing on $\alpha$ and $\beta$, the likelihood ratio test is oversized
and the gradient test becomes conservative.
However, the best performing test is the gradient test.
For example, when $n=20$, d.o.c.~= 10\%, $\gamma = 10\%$ and $\alpha = 1.8$,
the null rejection rates of the tests were
12.44\% ($LR_{(\alpha)}$), 9.77\% ($S_{T(\alpha)}$) and 9.83\% ($S_{T(\alpha)}^*$) for
testing $\mathcal{H}_{00}:\alpha=1.8$, and 11.90\% ($LR_{(\beta)}$) and 9.09\% ($S_{T(\beta)}$)
for testing $\mathcal{H}_{01}:\beta=1$.

A natural question at this point is why the Wald and the Rao score tests were not
included in the previous simulation experiments. Recall that the Wald and the score statistics
involve the Fisher information matrix, which cannot be obtained for the
$\mathcal{BS}$ distribution under type-II censoring. A common practice in such a case is
to use the observed information in the place of the expected information.
We followed this approach and ran various simulation experiments including the Wald and
the score tests. Our general conclusion is that these tests cannot be recommended for 
the following reasons. First, the Wald and score tests were markedly oversized.
For example, for $n=20$, $\gamma=10\%$ and $\alpha=0.75$,
the null rejection rates of the Wald and score tests for testing the null hypothesis
$\mathcal{H}_{01}:\beta=1$ are, respectively,  13.79\% and 15.61\% for d.o.c.~= 10\%,
16.73\% and 15.44\% for d.o.c.~= 30\% and 19.02\% and 13.70\% for d.o.c.~= 50\%.
Second, in our simulations the inverse of the observed information matrix frequently
produced negative standard errors in censored samples. 


In summary, the best performing test for $\alpha$ is the adjusted gradient test,
i.e.~the one that uses the bias-corrected estimator of $\alpha$. As far as hypothesis
testing on $\beta$ is concerned, the gradient test performs better than the likelihood
ratio test and should be preferred.


\section{Applications}\label{applications}

In this section, we shall illustrate the use of the likelihood ratio
and gradient tests in three real data sets. The first data set,
collected by members of the Instrument Development Unit of the Physical Research
Staff, Boeing Airplane Company, are the number of cycles-to-failure
of 101 strips of 6061-T6 aluminum sheeting, cut
parallel to the direction of rolling. Each had been subjected
to periodic loading with a frequency of 18 cycles per second,
and maximum stress of 31 kpsi. The data were taken from Table 8 of \cite{LCNV07}.
Suppose we are interested in testing $\mathcal{H}_{00}:\alpha=0.15$ against
$\mathcal{H}_{10}:\alpha\neq 0.15$ and $\mathcal{H}_{01}:\beta=125$
against $\mathcal{H}_{11}:\beta\neq 125$. These hypotheses
were considered in \cite{Cysneiros-et-al-2008}. Assuming
different censoring proportions, the observed values
of the different test statistics and the corresponding
$p$-values are given in Table~\ref{dataset1_tests}.
Note that for complete data without censoring ($m=101$)
and when $m=80$, at the 5\% significance level,
the likelihood ratio test does not reject the null
hypothesis $\mathcal{H}_{00}$ unlike the gradient tests.
For the other values of $m$, the likelihood ratio and the gradient tests lead to
the same conclusion. In all cases, the gradient tests provide stronger evidence
against the null hypothesis (smaller $p$-values than the likelihood
test). For testing the null hypothesis $\mathcal{H}_{01}$, the
same decision is reached by the likelihood ratio
and gradient tests, i.e.~the null hypothesis $\mathcal{H}_{01}$
is rejected at any usual significance level.
We also note that as the censoring proportions increase,
the $p$-values decrease.
\begin{table}[!htp]
\begin{center}
{\footnotesize
\caption{Test statistics ($p$-values between parentheses);
aluminum fatigue data.}\label{dataset1_tests}
\begin{tabular}{ccc c ccc}\hline
 &  \multicolumn{3}{c}{Inference on $\alpha$} && \multicolumn{2}{c}{Inference on $\beta$}\\ \cline{2-4}\cline{6-7}
$m$ & $LR_{(\alpha)}$ & $S_{T(\alpha)}$ & $S_{T(\alpha)}^*$ && $LR_{(\beta)}$ & $S_{T(\beta)}$ \\ \cline{1-4}\cline{6-7}
101 & 3.5771\,(0.0586) & 3.9841\,(0.0459) & 4.3171\,(0.0377) && 9.4279\,(0.0021) & 9.2402\,(0.0024) \\
 95 & 2.8573\,(0.0910) & 3.1598\,(0.0755) & 3.4821\,(0.0620) && 9.4250\,(0.0021) & 9.2582\,(0.0023) \\
 90 & 3.0826\,(0.0791) & 3.4342\,(0.0639) & 3.7969\,(0.0513) && 9.5167\,(0.0020) & 9.3800\,(0.0022) \\
 80 & 3.8361\,(0.0502) & 4.3641\,(0.0367) & 4.8300\,(0.0280) && 9.8999\,(0.0017) & 9.8573\,(0.0017) \\
 70 & 2.8684\,(0.0903) & 3.2360\,(0.0720) & 3.6615\,(0.0557) && 9.4412\,(0.0021) & 9.5939\,(0.0020) \\
 60 & 4.5172\,(0.0336) & 5.3218\,(0.0211) & 5.9240\,(0.0149) &&10.8407\,(0.0010) &11.3070\,(0.0008) \\
 50 & 4.0608\,(0.0439) & 4.8212\,(0.0281) & 5.4073\,(0.0201) &&10.3798\,(0.0013) &11.3522\,(0.0008) \\
 40 & 8.8234\,(0.0030) &11.6943\,(0.0006) &12.7236\,(0.0004) &&15.3808\,(0.0001) &18.0787\,(0.0000) \\\hline
\end{tabular}
}
\end{center}
\end{table}

As a second application, we consider the data provided by \cite{McColl1974} on the
the lifetime, in hours, of 10 sustainers of a certain type.
The data, used by \cite{Cohen-et-al-1984} as an
illustration of the three-parameter Weibull distribution,
can be found at Table 5 of \cite{LCNV07}.
We wish to test $\mathcal{H}_{00}:\alpha=0.21$ against
$\mathcal{H}_{10}:\alpha\neq 0.21$ and $\mathcal{H}_{01}:\beta=180$
against $\mathcal{H}_{11}:\beta\neq 180$. These hypotheses were
considered in \cite{Cysneiros-et-al-2008}. Under
different censoring proportions, the observed values
of the different test statistics and the corresponding
$p$-values are given in Table~\ref{dataset2_tests}.
Note that for complete data (without censoring)
the likelihood ratio test does not reject the null
hypothesis $\mathcal{H}_{00}$ at any usual significance level, whereas the original and
adjusted gradient tests reject the null hypothesis $\mathcal{H}_{00}$ at the 10\%
and 5\% significance levels, respectively.
Except for the case where $m=10$ (no censoring), the likelihood
and gradient tests lead to the same conclusion.
For testing the null hypothesis $\mathcal{H}_{01}$,
unless $m=7$, the same decision is reached by the likelihood ratio
and gradient tests, i.e.~the null hypothesis
is rejected at the 10\% significance level
for $m=8, 9$ and 10, and it not rejected
at any usual significance level for $m=6$.
\begin{table}[!htp]
\begin{center}
{\footnotesize
\caption{Test statistics ($p$-values between parentheses); sustainers data.}\label{dataset2_tests}
\begin{tabular}{ccc c ccc}\hline
 &  \multicolumn{3}{c}{Inference on $\alpha$} && \multicolumn{2}{c}{Inference on $\beta$}\\ \cline{2-4}\cline{6-7}
$m$ & $LR_{(\alpha)}$ & $S_{T(\alpha)}$ & $S_{T(\alpha)}^*$ && $LR_{(\beta)}$ & $S_{T(\beta)}$ \\ \cline{1-4}\cline{6-7}
10 & 2.1646\,(0.1412) & 2.7944\,(0.0946) & 4.0043\,(0.0454)   && 2.9417\,(0.0863) & 2.7580\,(0.0968) \\
 9 & 0.0770\,(0.7814) & 0.0728\,(0.7873) & 0.0000\,(1.0000) && 3.2449\,(0.0716) & 2.9248\,(0.0872) \\
 8 & 0.3307\,(0.5653) & 0.2911\,(0.5895) & 0.0000\,(1.0000) && 3.1616\,(0.0754) & 2.8499\,(0.0914) \\
 7 & 0.6732\,(0.4119) & 0.5514\,(0.4578) & 0.1472\,(0.7013)   && 2.9510\,(0.0858) & 2.7036\,(0.1001) \\
 6 & 0.8471\,(0.3574) & 0.6620\,(0.4159) & 0.2234\,(0.6365)   && 2.6797\,(0.1016) & 2.5463\,(0.1106) \\\hline
\end{tabular}
}
\end{center}
\end{table}

Finally, our third example uses the data given by \cite{Gupta1952}, which are
the survival times (in days) of the first seven of a sample of ten mice after being
inoculated with a culture of tuberculosis. The observations are: 41, 44, 46, 54, 55, 58
and 60. Suppose we are interested in testing $\mathcal{H}_{00}:\alpha=0.1$ against
$\mathcal{H}_{10}:\alpha\neq 0.1$ and $\mathcal{H}_{01}:\beta=54$
against $\mathcal{H}_{11}:\beta\neq 54$.
We have $LR_{(\alpha)} = 1.7607$ ($p$-value = 0.1845),
$S_{T(\alpha)}=2.3152$ ($p$-value = 0.1281),
$S_{T(\alpha)}^*=3.7346$ ($p$-value = 0.0533),
$LR_{(\beta)} = 1.3710$ ($p$-value = 0.2416) and
$S_{T(\beta)}=1.2054$ ($p$-value = 0.2722).
Note that the likelihood ratio and original gradient tests do
not reject the null hypothesis $\mathcal{H}_{00}$
at any usual significance level, whereas the
adjusted gradient test rejects the null hypothesis $\mathcal{H}_{00}$ at the 10\%
level. Additionally, the null hypothesis $\mathcal{H}_{01}$ is not rejected
at any usual significance level based on the statistics $LR_{(\beta)}$ and
$S_{T(\beta)}$.

\section{Concluding remarks}\label{conclusions}

The $\mathcal{BS}$ distribution is becoming increasingly popular
in lifetime analyses and reliability studies.
In this paper, we dealt with the issue of performing hypothesis testing
concerning the parameters of this distribution under
type-II right censored samples. We considered the
likelihood ratio and a recently proposed test, the gradient test.
As evidenced by our simulation results, in general the likelihood ratio
test is oversized (liberal), whereas the gradient test
is undersized (conservative) and less size distorted than the likelihood
ratio test. We also considered an adjusted gradient statistic
for testing hypotheses on the parameter $\alpha$
whose null distribution is more accurately approximated
by the limiting null distribution than that of the
original gradient test statistic.  Overall, our numerical results favor the
gradient tests for testing hypotheses on the parameters $\alpha$
and $\beta$. The results in this paper are encouraging and motivate further work. 
We plan to investigate the performance of the gradient test as compared to that of 
the likelihood ratio test in the Birnbaum--Saunders regression model under various censoring
mechanisms.


\end{document}